\documentclass[conference]{IEEEtran}
\IEEEoverridecommandlockouts

\usepackage{cite}
\usepackage{amsmath,amssymb,amsfonts}
\usepackage{algorithmic}
\usepackage{graphicx}
\usepackage{textcomp}
\usepackage{xcolor}

\usepackage{bm}
\usepackage{url}
\usepackage{subfig}
\usepackage{booktabs}
\usepackage{pifont}
\newcommand{\cmark}{\ding{51}} 
\newcommand{\xmark}{\ding{55}}

\def\BibTeX{{\rm B\kern-.05em{\sc i\kern-.025em b}\kern-.08em
    T\kern-.1667em\lower.7ex\hbox{E}\kern-.125emX}}
\begin{document}

\title{Research Paper Recommender System by Considering Users' Information Seeking Behaviors
\thanks{The research was supported by Faculty of Strategy Headquarters, Department of Comprehensive Human Science, University of Tsukuba.}
}
\author{\IEEEauthorblockN{Zhelin Xu}
\IEEEauthorblockA{\textit{Academic Service Office for the Library,} \\ 
\textit{Information and Media Sciences Area} \\
\textit{University of Tsukuba}\\
Tsukuba, Ibraki, Japan \\
zhelin@ce.slis.tsukuba.ac.jp}
\and
\IEEEauthorblockN{Shuhei Yamamoto}
\IEEEauthorblockA{\textit{Institute of Library,}\\
\textit{Information and Media Science} \\
\textit{University of Tsukuba}\\
Tsukuba, Ibraki, Japan \\
syamamoto@slis.tsukuba.ac.jp	}
\and
\IEEEauthorblockN{Hideo Joho}
\IEEEauthorblockA{\textit{Institute of Library,}\\
\textit{Information and Media Science} \\
\textit{University of Tsukuba}\\
Tsukuba, Ibraki, Japan \\
hideo@slis.tsukuba.ac.jp		}
}

\maketitle

\begin{abstract}
With the rapid growth of scientific publications, researchers need to spend more time and effort searching for papers that align with their research interests. 
To address this challenge, paper recommendation systems have been developed to help researchers in effectively identifying relevant paper. 
One of the leading approaches to paper recommendation is content-based filtering method. 
Traditional content-based filtering methods recommend relevant papers to users based on the overall similarity of papers. 
However, these approaches do not take into account the information seeking behaviors that users commonly employ when searching for literature. 
Such behaviors include not only evaluating the overall similarity among papers, but also focusing on specific sections, such as the method section, to ensure that the approach aligns with the user's interests. 
In this paper, we propose a content-based filtering recommendation method that takes this information seeking behavior into account. Specifically, in addition to considering the overall content of a paper, our approach also takes into account three specific sections (background, method, and results) and assigns weights to them to better reflect user preferences.
We conduct offline evaluations on the publicly available DBLP dataset, and the results demonstrate that the proposed method outperforms six baseline methods in terms of precision, recall, F1-score, MRR, and MAP.

\end{abstract}


\section{Introduction}\label{sec:introduction}
Finding relevant research papers in a specific field is a fundamental task for researchers \cite{b1}. 
Reading literature related to their interests not only grants them access to valuable references but also keep them informed of the latest techniques \cite{b2}. Moreover, it helps them pinpoint novel aspects of their own work \cite{b3}.
However, due to the exponentially increasing number of papers published each year, 
researchers often spend significant time and effort locating publications that truly match their interests. 

To address this challenge of information overload, 
researchers often rely on academic search engines (e.g., Google Scholar\footnote{https://scholar.google.com/} or ACM Digital Library\footnote{https://dl.acm.org/}, etc.) and then review the retrieved articles to determine their relevance \cite{b32}.  
However, this process demands substantial expertise, as selecting effective keywords requires deep familiarity with the field \cite{b2}. 
For novice researchers, such as students who lack sufficient domain knowledge, defining suitable search terms can be particularly difficult \cite{b4,b5}. 
Consequently, they often need to repeatedly refine keywords to achieve satisfactory results \cite{b6}. 
Pre-trained large language model-based chatbot (e.g., ChatGPT, Perplexity AI) can also be utilized for literature searches, enhancing retrieval efficiency. 
However, these chatbots face two major challenges: (1) they may generate non-existent literature\cite{b43}; and (2) the effectiveness of retrieving relevant results depends on well-crafted prompts. However, developing effective prompting strategies remains a complex and challenging task\cite{b44}.

\begin{figure}[t]
\centerline{\includegraphics[width=0.5\textwidth]{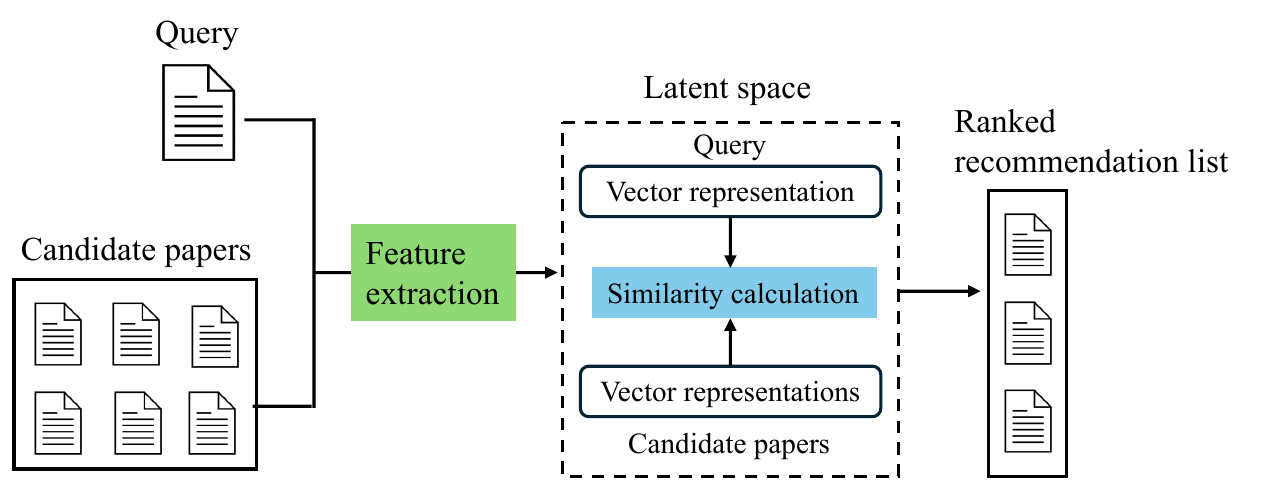}}
\caption{Overview of the paper recommendation task based on content-based filtering.}
\label{fig:RS_task}
\end{figure}

An alternative approach is to use recommender systems, which can rapidly identify valuable and relevant papers from large datasets \cite{b7}. 
A paper recommender system can be broadly defined as follows: given an input paper as query, it generates a ranked list of related articles\cite{b8}. 
Among various recommendation approaches, content-based filtering (CBF) is one of the most widely used techniques \cite{b9,b10}.
As shown in Fig. \ref{fig:RS_task},  CBF typically extracts textual features from the content of both the query and candidate papers, such as titles, abstracts, keywords, or the main body. 
These features are then map into a latent space to produce vector representations, which are used to compute semantic similarity. 
Based on the resulting similarity scores, a ranked list is generated, with the top-ranked papers are recommended to the user as relevant paper.
\setlength{\textfloatsep}{8pt}

Although numerous CBF-based methods for paper recommendation have been proposed \cite{b11,b12,b13,b14}, 
these approaches only focus on overall similarity between papers, 
overlooking users' information seeking behaviors in literature exploration. 
A study investigating the information seeking behaviors of master’s students revealed that they typically begin by examining a paper’s abstract and then proceed to review specific sections of the article before determining its relevance \cite{b15}. 
This finding implies that users not only consider the overall similarity between papers but also examine specific sections to confirm whether a paper aligns with their interests.
However, most CBF-based methods do not incorporate the importance of these specific sections and may therefore fail to satisfy user needs. 
For example, consider a user searching for e-commerce recommendation research that uses recurrent neural networks (RNN). This user might examine both the paper’s overall relevance and its methodology section to determine whether an RNN-based approach is employed.
In contrast, traditional CBF-based methods rely solely on global similarity, 
which might result in recommending a candidate paper that utilizes collaborative filtering techniques simply because it belongs to the same e-commerce recommendation domain and uses a similar offline evaluation approach as the query paper. 
Since the user is specifically interested in RNN-based recommendation methods, this recommendation would not align with their needs.

In this paper, we propose a new method for recommending relevant research papers to novice researchers. 
Our approach learns a new paper representation by combining the paper's overall content with three weighted sections (background, method and results). 
Due to its alignment with common information seeking behaviors in literature exploration, this method aims to provide more accurate results. We validate our method through an offline experiment on the public DBLP dataset. 
The results show that our method achieves a recall@5 of 0.8125 and a MAP of 0.8081. 
These scores represent a 3.1\% and 3.2\% improvement, respectively, compared to the previous best results. 
These findings demonstrate the effectiveness of our approach.

Our contributions are as follows: 
\begin{itemize}
    \item Considering users' information seeking behavior in literature exploration, we propose a new paper recommendation method that takes into account both overall content of a paper and the information in three weighted specific sections extracted from the paper. 
    \item Our method achieves state-of-the-art performance on the public DBLP dataset for paper recommendation task. 
    \item We provide a detailed analysis of the proposed method, and further present a case study to demonstrate the paper recommended to the user by our approach.
\end{itemize}

\section{Related Work}
Three types of approaches are commonly used in paper recommendation systems. 
Graph-based approaches recommend papers based on network structures. Collaborative filtering approaches consider the preferences of users with similar interests. 
CBF approaches focus on the content similarity between papers. 
In the following paragraphs, we will describe each type of approach. 

\textbf{Graph-based Approaches.} 
Several studies have proposed utilizing graph-based methods for paper recommendation \cite{b19,b20}. 
These methods typically construct a heterogeneous network and exploit its topological structure to generate recommendations \cite{b16}. 
Specifically, the approach generally involves three steps: 
(1) building a network where nodes represent entities such as authors, venues, or papers,
and edges capture relationships like authorship or citation, 
(2) embedding nodes into vector using techniques such as DeepWalk \cite{b17} or node2vec \cite{b18}, 
and (3) generating recommendations based on similarity between these node embeddings. 
However, these approaches often overlook the textual content of papers \cite{b21}, potentially leading to recommendations that lack content relevance to the target papers. 
For novice researchers, such as students, content similarity is especially important. 
Due to their limited understanding of the field, providing recommendations with closely related content can guide them in conducting deeper explorations of topics of interest and foster a better understanding of this research field. 

\textbf{Collaborative filtering Approaches.} 
Collaborative filtering (CF) methods are based on the assumption that users who share interests in certain papers are likely to have similar preferences for other papers \cite{b22}. 
Traditionally, CF approaches use explicit feedback, such as paper ratings on platforms like CiteULike to capture user interests \cite{b23}. 
However, new users often lack a sufficient rating history, and most users tend to provide limited explicit feedback, posing significant challenges \cite{b24}. 
This situation leads to the cold-start problem, where the system lacks sufficient data to identify users with similar preferences. 
To address this problem, some studies utilize implicit feedback to infer user interests based on system interactions, such as downloading papers \cite{b25}. 
Nevertheless, 
implicit feedback may not perfectly mirror real-world user behavior \cite{b31}, 
and CF methods also do not consider the textual content in papers. 

\textbf{Content-based filtering Approaches.} As mentioned in section \ref{sec:introduction}, CBF methods typically extract textual features from various parts of a paper to calculate the semantic similarity. 
The features are derived from the title, abstract, keywords, main body and venue \cite{b11,b12,b13,b14,b26,b27}.
Furthermore, some approaches also incorporate supplementary information such as paper tags or popularity metrics to refine similarity calculations \cite{b28}. 
Unlike our proposed method, most existing CBF techniques focus on computing an overall similarity score between papers without considering the different importance of specific sections (e.g., background, method, results). 
A few studies have recognized that different parts of a paper should be assigned different weights. 
For example, \cite{b29} extracts the top 10 important phrases from the title, abstract, main body to represent the paper. 
In contrast, our proposed method differs in two key ways: (1) our approach uses section-level information, 
and (2) we do not exclude sections that might appear less important, and we also consider the overall similarity of papers. 
Additionally, \cite{b30} assigns weights to the abstract, author and venue, respectively. Unlike our proposed method, it does not account for varying importance within different sections (e.g., background, methods, results) of a paper.

\section{Method}
\subsection{Overview}
\begin{figure*}[htb!]
\centerline{\includegraphics[width=0.95\textwidth]{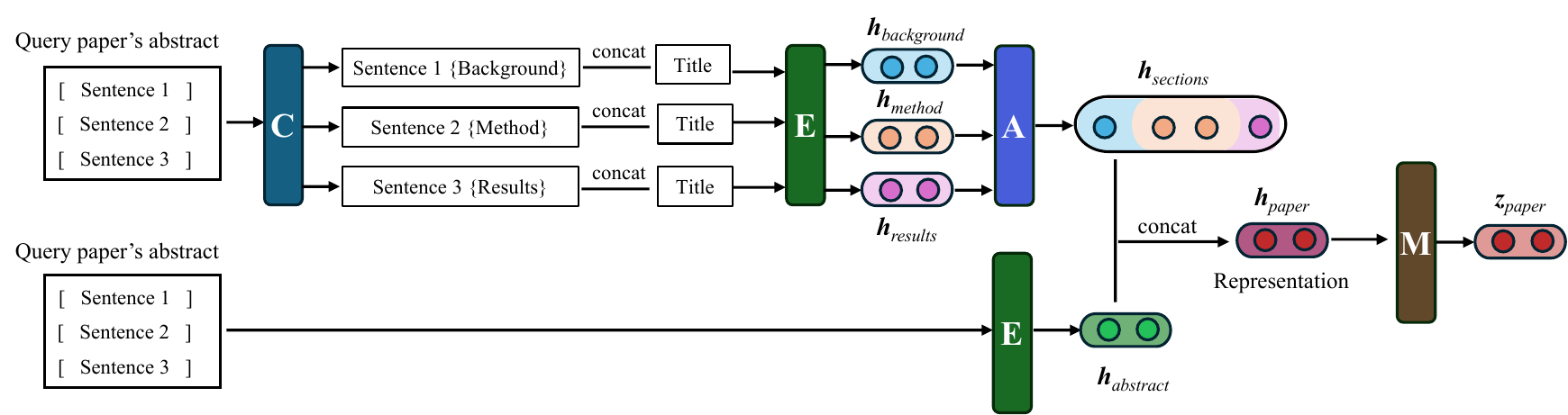}}
\caption{Overview of Proposed Method. C: Classification model, this model is used to extract three specific sections from the query paper's abstract. Title: the title of the query paper, which is then appended to each extracted section. E: Embedding model, this model is used to encode each section or the full abstract into vector representations. A: Attention model, which assigns different weights to the extracted section embeddings. 
M: An non-linear MLP model.}
\label{fig:proposed_method}
\end{figure*}

Fig. \ref{fig:proposed_method} presents an overview of the proposed method. 
Instead of relying on manually defined query keywords, our approach selects a paper preferred by the user as the query.
This design makes it easier for novice researchers to get started. 
Similar to the query-by-example approach, in our model, the input query is a paper selected by the user and the goal is to recommend relevant papers from a large number of candidate papers.
In particular, the model processes this query and identifies a set of independent candidate papers by computing their similarity to the query. These candidate papers are then ranked based on the similarity, forming a ranked list. The output consists of the top-ranked papers from this list, which are then recommended to the user, respectively. Due to copyright restrictions, the full text of many research papers is not publicly accessible, making it unavailable for recommendation tasks \cite{b33}. 
Reference \cite{b9} has highlighted that the abstract provides the most accurate description of the authors' work, while the title serves as a concise summary of the article. 
Based on these findings, our method utilizes the abstract and title to represent each paper.

Since we consider users' information seeking behaviors, 
the representation of the paper is designed to integrate both its section-level information and overall content. Specifically, section-level information is captured in $\bm{h}_{\text{sections}}$, while the overall content is represented by $\bm{h}_{\text{abstract}}$. 
Each sentence in the abstract is processed by a \textbf{C}lassification model, an \textbf{E}mbedding model, and an \textbf{A}ttention model to generate the vector $\bm{h}_{\text{sections}}$. 
To enhance the representation of each section, query paper's title is concatenated with three sections (background, method, results) before embedding. 
As a result, $\bm{h}_{\text{sections}}$ includes both the section-level information and title information, with different weights assigned to each section. 
Meanwhile, the entire abstract is fed into the \textbf{E}mbedding model to produce a vector $\bm{h}_{\text{abstract}}$.
These two vectors are then concatenated to obtain $\bm{h}_{\text{paper}}$, which serves as a representation of the query paper. 
To further refine the quality of this representation, $\bm{h}_{\text{paper}}$ is input into a multi-layer perceptron (\textbf{M}LP) model to produce  $\bm{z}_{\text{paper}}$. 
This vector is then utilized to compute the loss during training. 
The following subsections provide a detailed explanation of the proposed method.

\subsection{Classification Model}\label{sec:classification}
We adopt the classification model proposed by \cite{b34} to categorize each sentence in an abstract. 
This is a BERT-based classification model and does not require additional complex architectural augmentations, such as conditional random fields.
The model defines five types of categories: background, method, results, objective, and other. 
When a sentence is input into the model, it produces a probability distribution over these five categories (e.g., [0.93, 0.03, 0.01, 0.02, 0.01]).
Since many papers are structured around the three categories of background, method, and results, these categories contain key information. 
Moreover, the abstract is highly likely to include these categories. 
Therefore, we focus on these three categories in our approach. Each sentence is assigned to the category for which it has the highest predicted probability. 
Note that each section representing a specific category can contain zero, one, or multiple sentences. 

\subsection{Embedding Model}
After classifying the sentences in the abstract, the paper’s title is incorporated into each section. 
Because one section may contain multiple sentences, methods designed for sentence-level or token-level tasks are less effective for section-level representation. 
To address this, we employ the SPECTER model \cite{b12} to embed each section. 
SPECTER is a SciBERT-based method and is designed to produce document-level embeddings for research papers by considering relationships among papers. 
As a result, three vectors $\bm{h}_{\text{background}}$, $\bm{h}_{\text{method}}$, and $\bm{h}_{\text{results}}$ are obtained, each vector contains the information from a specific category and the title. 
Additionally, the entire abstract is also processed by SPECTER to generate a vector $\bm{h}_{\text{abstract}}$, representing the abstract’s overall content.

\subsection{Attention Model}\label{sec:attention}
As mentioned in section \ref{sec:introduction}, an analysis of information seeking behaviors in literature exploration shows that researchers also pay attention to whether specific sections of a candidate paper match their needs. 
For example, if a user is looking for work on paper recommender systems that utilize the CBF approach, this indicates a specific interest in studies employing the CBF method. In this case, assigning a higher weight to the method section in the query paper may be necessary to address their needs.

Existing public datasets often lack detailed information about users' specific needs. Gathering such data through surveys would entail substantial costs, particularly for large-scale user studies. 
In addition, novice researchers such as undergraduate and graduate students, often have limited knowledge in their areas of interest. 
As a result, they may find it challenging to assign appropriate weights to each section of a paper by themselves. 
To address this issue, we use a multi-head attention model \cite{b35} to automatically estimate the weights of three specific sections. Sections with higher weights are considered more relevant to the user's needs. 

Specifically, vectors $\bm{h}_{\text{background}}$, $\bm{h}_{\text{method}}$, and $\bm{h}_{\text{results}}$ are first stacked and then fed into the attention model. 
The attention mechanism generates a set of refined vectors [$h_{\text{background}}^{'}$, $h_{\text{method}}^{'}$, $h_{\text{results}}^{'}$], where each vector contains weighted information from all three sections. 
Finally, these refined vectors are averaged to obtain $\bm{h}_{\text{sections}} = (h_{\text{background}}^{'}+h_{\text{method}}^{'}+h_{\text{results}}^{'})/3$. 


\subsection{Similarity Calculation}\label{sec:sim}
To align with users' information seeking behaviors, the paper representation should contain both its weighted section-level information and overall content. 
Additionally, we assume that $\bm{h}_{\text{sections}}$ and $\bm{h}_{\text{abstract}}$ may have different weights.
Thus, they are combined to calculate the paper representation using \eqref{eq:h_paper}:
\begin{equation}
\bm{h}_{\text{paper}}=\alpha \times \bm{h}_{\text{sections}} + (1-\alpha) \times \bm{h}_{\text{abstract}}
\label{eq:h_paper}
\end{equation}
$\alpha$ is a hyperparameter that define the weights assigned to each vector.
The resulting vector $\bm{h}_{\text{paper}}$ is used to measure the cosine similarity between papers, such as a query paper ($\text{paper}_i$) and a candidate paper ($\text{paper}_j$), as defined in \eqref{eq:cos}:
\begin{equation}
similarity(\text{paper}_i, \text{paper}_j) = \frac{\left(\bm{h}_{\text{paper}}^i\right)^T \bm{h}_{\text{paper}}^j}{\|\bm{h}_{\text{paper}}^i\| \cdot \|\bm{h}_{\text{paper}}^j\|}
\label{eq:cos}
\end{equation}

\subsection{MLP Model}\label{sec:mlp}
Previous research pointed out that applying a learnable non-linear transformation to representations before computing the loss can enhance the quality of the learned embeddings \cite{b36}. 
In this paper, we implement this approach by using a MLP model with one hidden layer to produce the vector $\bm{z}_{\text{paper}}$, as defined in \eqref{eq:z_paper}:
\begin{equation}
\bm{z}_{\text{paper}} = {W}^{(2)} \cdot ReLU\left({W}^{(1)} \cdot \bm{h}_{\text{paper}} + {b}^{(1)}\right) + {b}^{(2)}
\label{eq:z_paper}
\end{equation}
${W}^{(1)}$ and ${b}^{(1)}$ are the learnable weight matrix and bias term responsible for transforming the input representation into the hidden layer, 
while ${W}^{(2)}$ and ${b}^{(2)}$ refer to those used for the transformation from the hidden layer to the final output.
The ReLU activation introduces non-linearity. 
Finally, the output $\bm{z}_{\text{paper}}$ is used to compute the loss. Furthermore, as demonstrated in section \ref{sec:nonlinear}, the experimental results indicate that using $\bm{z}_{\text{paper}}$ instead of $\bm{h}_{\text{paper}}$ for loss calculation achieves better results.

\subsection{Pretraining Objective}\label{sec:pretraing_objective}
\begin{figure}[htb!]
\centerline{\includegraphics[width=0.35\textwidth]{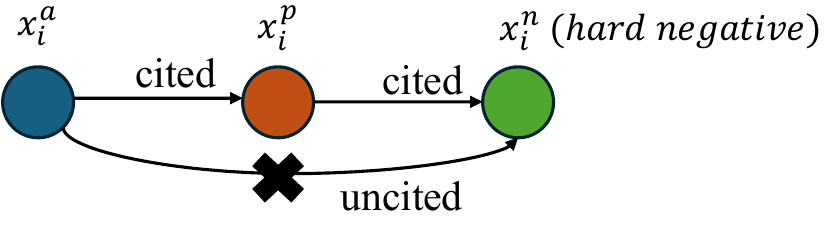}}
\caption{Positive and hard-negative sample selection for a query paper based on citation relationships.}
\label{fig:loss_samples}
\end{figure}

Our training objective is to minimize the distance between the query paper and its relevant papers, while maximizing the distance between the query paper and irrelevant papers. This design enhances the model’s ability to capture relevant relationships.
Therefore, we adopt a triplet loss function \cite{b37}, shown in \eqref{eq:tripletloss}:
\begin{equation}
L = \sum_{i=1}^{N}max\{d(x_i^a,x_i^p)-d(x_i^a,x_i^n)+m, 0\}
\label{eq:tripletloss}
\end{equation}
Here, as shown in Fig. \ref{fig:loss_samples}, $x_i^a$ is a query paper. 
$x_i^p$ represents a positive sample, defined as a relevant paper to the query paper such as one cited by the query paper \cite{b2}. 
$x_i^n$ is a negative sample. 
To improve the model’s performance, we adopt the approach proposed in \cite{b12} by increasing the difficulty of training. This approach defines two types of negative samples:
(1) hard negatives: if the query paper cites paper A, which in turn cites paper B, but the query paper does not cite paper B, then paper B (represented by the green node in the Fig. \ref{fig:loss_samples}) is considered a hard negative sample for query paper, 
and (2) random negatives: papers randomly sampled from the dataset serve as negative examples for the query paper. 
$N$ represents the total number of triples constructed from the dataset. 
$m$  is a margin hyperparameter, which we set to 1 in our experiments. 
The distance between the query paper and the positive (or negative) sample is computed using the L2 norm distance, as shown in \eqref{eq:L2_distance}:
\begin{equation}
d(x_i^a,x_i^p) = \|(f(x_i^a)-f(x_i^p))\|_2
\label{eq:L2_distance}
\end{equation}
The proposed method is represented by $f(x)$, which embeds each paper into a 786-dimensional vector $\bm{z}_{\text{paper}}$. 

\section{Experiment}
\subsection{Experimental Data}\label{sec:experiment data}
Offline evaluation measures model performance using a dataset and does not require user participation, making it both efficient and popular \cite{b2}. 
In this paper, we utilize the publicly available DBLP-Citation-network V1\footnote{\url{https://www.aminer.cn/citation}} dataset, which contains 63K papers. 
We use paper abstracts and titles to identify relevant papers. 
Specifically, papers in a query paper's reference list are regarded as its relevant papers.  
However, only a portion of these papers provide both abstracts and references. 
We therefore select papers that meet the following criteria as query papers: (1) the paper has both references and an abstract, and (2) its references also have abstracts. 
We then create three datasets from these query papers as described below and summarized in Table \ref{tabl:dataset}:

\begin{itemize}
    \item Train dataset: Contains 43K query papers. 
    Each query paper is associated with a candidate set that includes one positive sample (randomly chosen from its references), one hard negative sample (randomly chosen from its hard negative samples), and 11 random negative samples. This setup results in a total of 516K triples for loss computation. 
    \item Validation dataset: Includes about 4.2K query papers. The candidate set for each query paper contains an average of three positive samples and 100 random negative samples, where the number of negative samples follows the approach suggested in \cite{b41}. As a result of this setup, 1.26M triples are generated for validation. 
    \item Test dataset: Consists of 4.2K query papers. Similar to the validation dataset, the candidate set for each query paper consists of an average of three positive samples and 100 random negative samples.
\end{itemize}

\begin{table}[htbp]
\centering
\caption{Statistics of the evaluation datasets. }
\begin{tabular}{lccc}
\toprule
 & Train & Validation & Test  \\
\midrule
Query  & 43k & 4.2K & 4.2K \\
Positive & 43K  & 12.6K & 12.6K \\
Hard Negative & 43K  & 0 & 0   \\
Random Negative &  473K  & 420K & 420K \\
\bottomrule
\end{tabular}
\label{tabl:dataset}
\end{table}

After training the proposed model on the training dataset, our goal is to identify relevant papers (positive samples) from candidate set for each query paper in the test data. 
Note that each query paper is used in only one of the three datasets to ensure no overlap.
In addition, for the validation and test datasets, none of the candidate items for a given query paper are present in the train dataset.

\subsection{Training Details}
We initialize the embedding model with the pretrained SPECTER \cite{b12} weights, while the attention and MLP models are implemented in PyTorch\footnote{https://pytorch.org/docs/stable/nn.html}. 
We then continue training all model parameters according to our training objective (as shown in \eqref{eq:tripletloss}). 
Note that only the last four layers of the embedding model are updated. 

Hyperparameter tuning is guided by performance on the validation dataset. 
For optimization, we adopt the Adam optimizer with a weight decay of 1e-4, setting the learning rate to 2e-5 for the embedding model and 5e-5 for other models. 
In the attention model, the number of heads is set to 4. 
We set $\alpha$ to 0.3, meaning that $\bm{h}_{\text{sections}}$ and $\bm{h}_{\text{abstract}}$ are assigned weights of $\alpha$ = 0.3 and 1 - $\alpha$ = 0.7, respectively. 
The model is trained on a single RTX 3080 Ti GPU with 12GB of memory for 4 epochs, using a batch size of 3, which is the maximum capacity that fits within our GPU memory.

\subsection{Baseline Methods}
We compare our approach against the following baseline models:
\begin{itemize}
    \item SPECTER \cite{b12}: A state-of-the-art method for learning scientific document representations by considering inter-document relatedness. 
    We do not fine-tune SPECTER, as the original paper states that its pretrained model does not require any task-specific fine-tuning.
    \item SimCSE \cite{b38}: A contrastive learning method for learning sentence embeddings. We train a SimCSE model from scratch on our training dataset and also fine-tune a pretrained SimCSE model on the same dataset. 
    \item BERT \cite{b39}: A pretrained transformer-based language model. We use both BERT-base\footnote{https://huggingface.co/google-bert/bert-base-uncased} and BERT-large\footnote{https://huggingface.co/google-bert/bert-large-uncased} as baseline methods.
    \item Doc2Vec \cite{b40}: An unsupervised approach for learning document embedding. 
    Following the hyperparameter settings in \cite{b12}, we train Doc2Vec on our own training dataset.
\end{itemize}

\subsection{Evaluation Metrics}
As shown in Fig. \ref{fig:RS_task}, our method generates a ranked list based on the similarity between papers and selects the top-ranked papers to recommend to the user. 
In this paper, we aim to evaluate our method in terms of the following three aspects: 
(1) whether all relevant papers achieve high positions in the ranking list, (2) whether the first relevant paper appear in the top position, and (3) whether the ranking list includes a large number of relevant papers. 
To evaluate these aspects, we conduct an offline experiment and evaluate the proposed method using the following metrics: Mean Average Precision (MAP), Mean Reciprocal Rank (MRR) and recall@N. 
In addition, we also employ commonly used ranking metrics such as precision@N and F1-score@N to assess the performance of the proposed method.

For instance, MAP is defined in \eqref{eq:MAP}:
\begin{equation}
MAP = \frac{1}{|Q|} \sum_{q \in Q} AP(q)
\label{eq:MAP}
\end{equation}
where $Q$ represents the number of query papers in each dataset. 
$AP(q)$ is computed using \eqref{eq:AP}:
\begin{equation}
AP(q) = \sum_{k=1}^{N} \frac{precision@k \cdot rel(k)}{|Relevant\ Papers|}
\label{eq:AP}
\end{equation}
Here, $N$ is the total number of candidate papers. 
$rel(k) = 1$ if the paper at rank $k$ is a relevant paper for the query, otherwise, $rel(k) = 0$. 
$|Relevant\ Papers|$ represents the total number of relevant papers for the query. 
Since the DBLP dataset does not include human-annotated relevance labels or user interaction data, many studies treat a query paper’s references as its relevant papers \cite{b2}. We follow this approach and consider the references of each query paper to be its relevant papers, corresponding to the positive samples discussed in section \ref{sec:pretraing_objective}. 

MRR is defined in \eqref{eq:MRR}:
\begin{equation}
MRR = \frac{1}{|Q|} \sum_{i=1}^{|Q|} \frac{1}{rank_i}
\label{eq:MRR}
\end{equation}
$rank_i$ represents the position of the first relevant paper in the ranked list for the $i$-th query. 


\subsection{Results}
Table \ref{tabl:MRR_MAP} shows the MAP and MRR results on the paper recommendation task. The results indicate that the proposed method consistently outperforms all baseline methods. 
In particular, the MAP score reaches 0.8081, representing a 3.2\% improvement over the next best baseline, SPECTER. 


\begin{table}[htbp]
\centering
\caption{MAP and MRR results on the paper recommendation task. SimCSE\_A refers to the model trained from scratch on the DBLP dataset, while SimCSE\_B refers to the pretrained model fine-tuned on the DBLP dataset.}
\renewcommand{\arraystretch}{1.4} 
\setlength{\tabcolsep}{13pt} 
\begin{tabular}{lcc}
\toprule
\textbf{Method} & \multicolumn{2}{c}{\textbf{Metric}} \\
\cmidrule(lr){2-3}
 & MAP & MRR \\
\midrule
Doc2Vec~\cite{b40}  & 0.5775 & 0.7150  \\
BERT-based~\cite{b39} & 0.1272 & 0.2165 \\
BERT-large~\cite{b39} & 0.0929 & 0.1492 \\
SimCSE\_A~\cite{b38} & 0.5569 & 0.7158  \\
SimCSE\_B~\cite{b38} & 0.5490 & 0.7023  \\
SPECTER~\cite{b12} & 0.7829 & 0.8693  \\
\midrule
\textbf{Proposed method} & \textbf{0.8081} & \textbf{0.8860}  \\
\bottomrule
\end{tabular}
\label{tabl:MRR_MAP}
\end{table}

\begin{figure*}[htbp]
    \centering
    \subfloat[Precision results of seven methods.]{
        \includegraphics[width=0.30\textwidth]{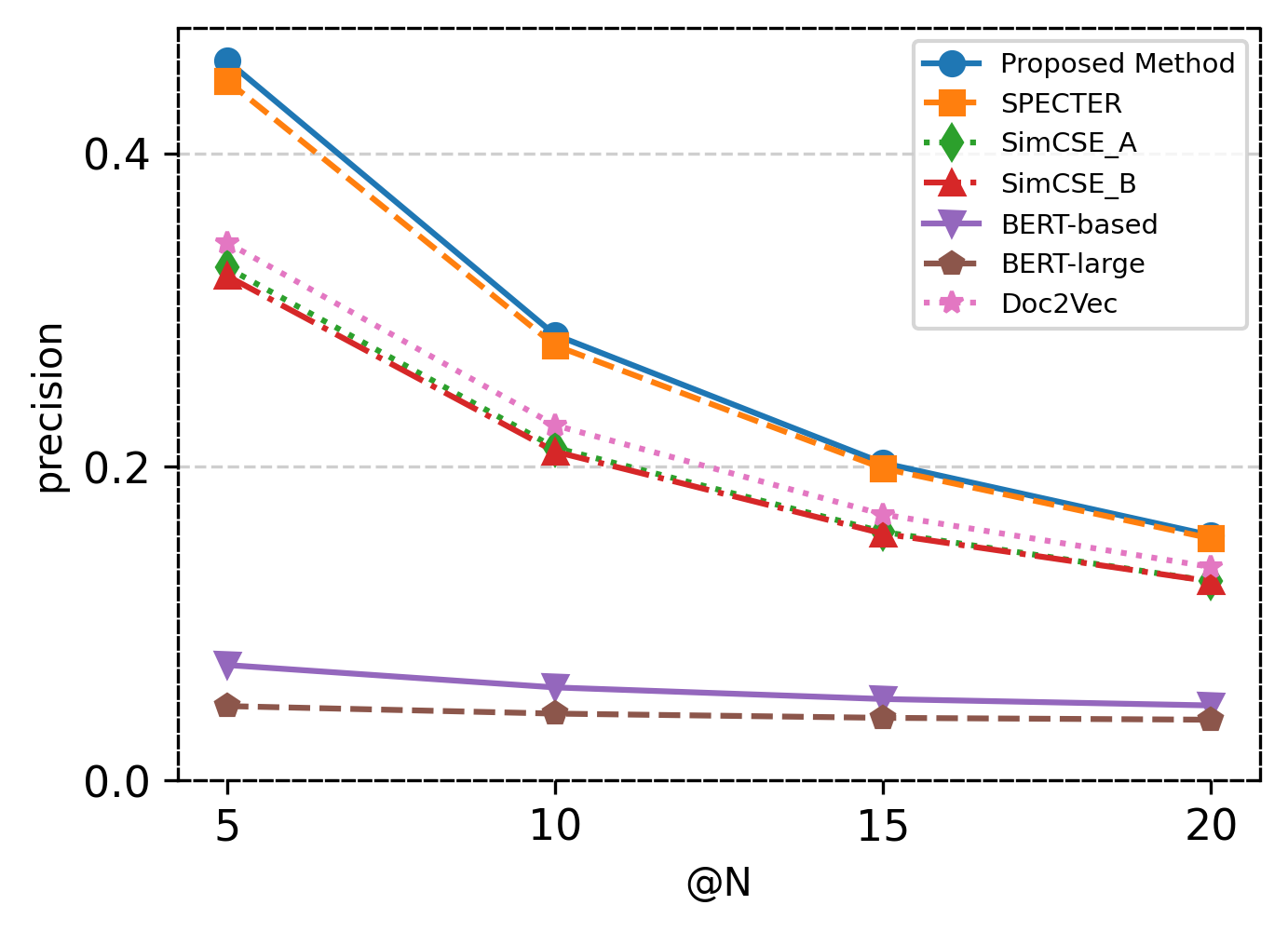}
    }
    \hspace{0.01\textwidth}
    \subfloat[Recall results of seven methods.]{
        \includegraphics[width=0.30\textwidth]{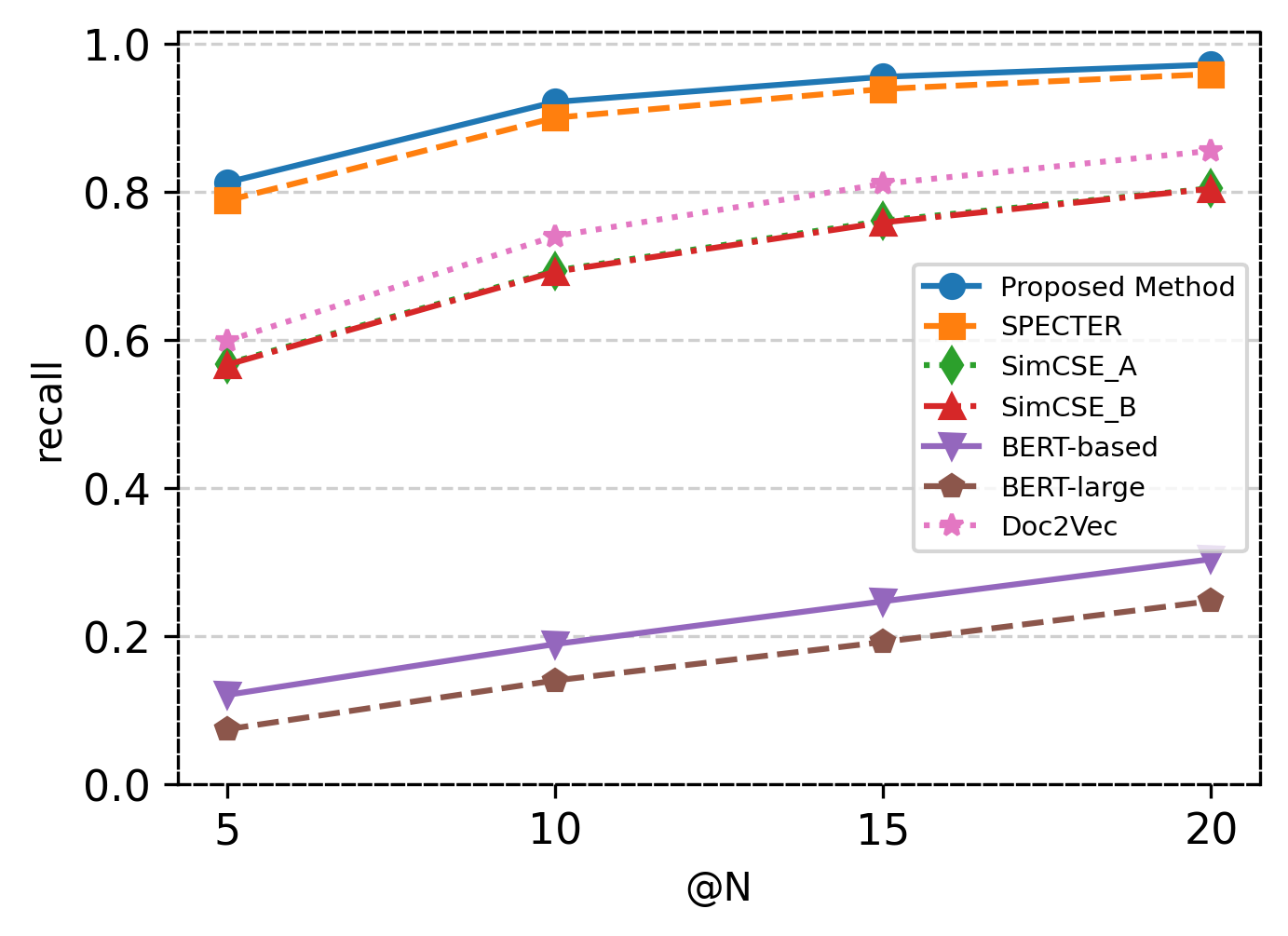}
    }
    \hspace{0.01\textwidth}
    \subfloat[F1-score results of seven methods.]{
        \includegraphics[width=0.30\textwidth]{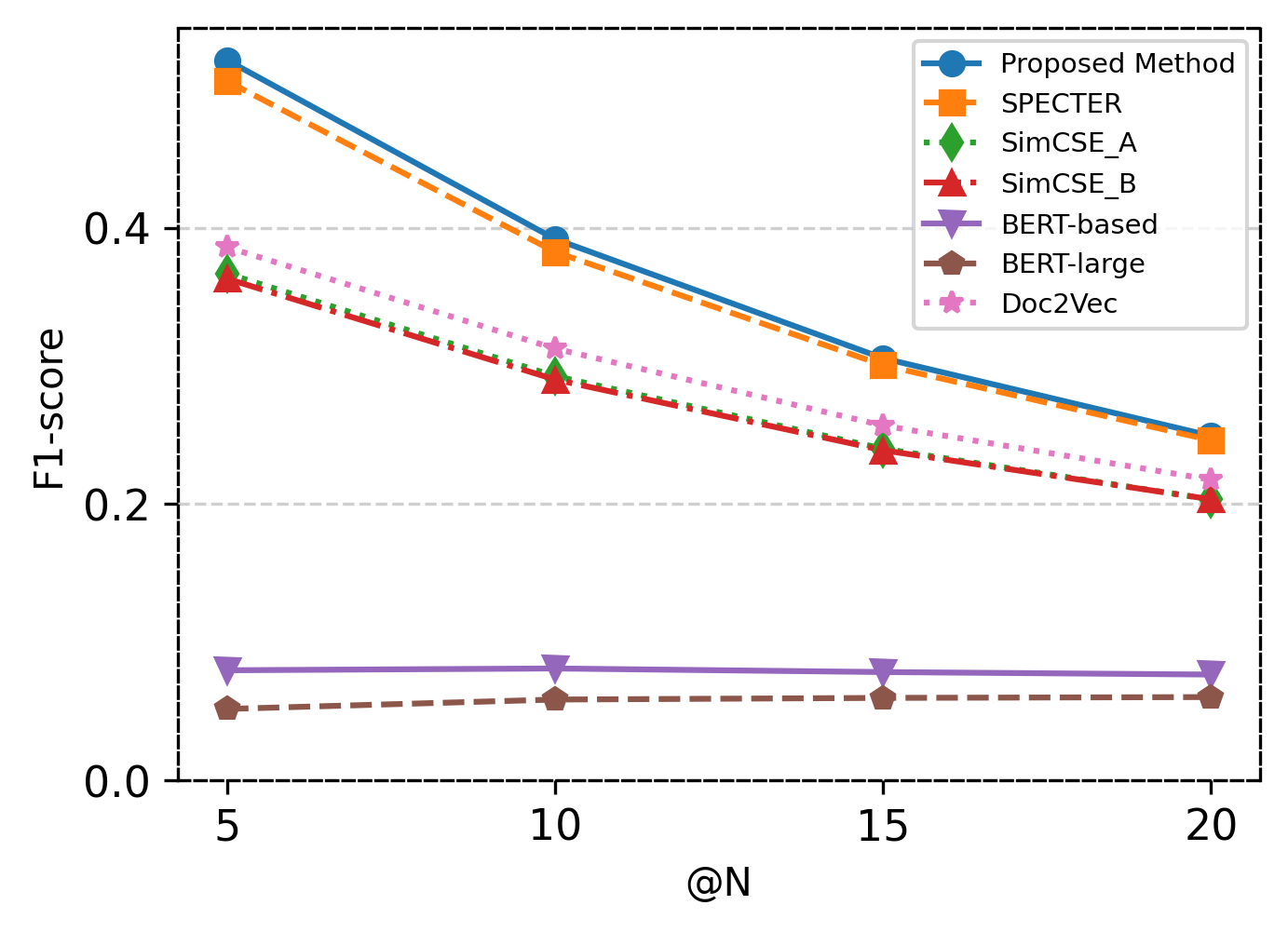}
    }
    \caption{The evaluation results of precision@N, recall@N and F1-score@N on the paper recommendation task.}
    \label{fig:PRF}
\end{figure*}

Fig. \ref{fig:PRF} presents the evaluation results of precision@N, recall@N and F1-score@N on the paper recommendation task. We observe that the proposed method outperforms all baseline methods. 
Additionally, \cite{b42} suggests that the optimal number of recommended papers lies between 5 and 6. 
Based on this insight, we confirm that the proposed method achieves
precision@5 = 0.4591 and recall@5 = 0.8125, representing 2.7\% and 3.1\% improvements, respectively, over the second-best baseline, SPECTER. 
Furthermore, the proposed method achieved 0.9717 in term of recall@20, indicating that when recommending the top 20 papers, it can effectively retrieve almost all relevant papers associated with the query paper. 


These experimental results validate the effectiveness of the proposed method in enhancing the performance of paper recommendation. 
Compared to baseline methods, by considering users' information seeking behavior, our approach creates a novel paper representation that incorporate both the overall information of a paper and its weighted section-level information. 
Since this representation better satisfies users' needs, the proposed method yields more accurate recommendations, thereby allowing more relevant papers to appear higher in the recommendation list.

\section{Analysis}
\subsection{Ablation Study}
We analyze how different components in the proposed method affect performance with the results presented in Table \ref{tabl:Ablation}. 
The first design decision in our model is to use a classification model to extract three specific sections which are background, method and results. 
In pattern \ding{172}, each paper's sections are encoded into vectors $\bm{\tilde{h}}_{\text{background}}$, $\bm{\tilde{h}}_{\text{method}}$ and $\bm{\tilde{h}}_{\text{results}}$ using the embedding model. 
The paper representation is then obtained by averaging these vectors. The similarity between papers is computed based on this representation. 
We observe that pattern \ding{172} results in a substantial decrease in performance, indicating that ignoring the differences among sections may negatively impact recommendation performance.

Pattern \ding{173} extends pattern \ding{172} by incorporating the paper's title into each section before embedding. 
The results show that adding the title improves performance. 
This suggests that the title provides valuable contextual information, which is beneficial for the paper recommendation task.

Pattern \ding{174} further improves pattern \ding{173} by introducing an attention model. 
After the embedding process, this model assigns adaptive weights to each section and then generates $\bm{h}_{\text{sections}}$. 
In this pattern, the final paper representation is defined as $\bm{h}_{\text{paper}} = \bm{h}_{\text{sections}}$. 
The results indicate that pattern \ding{174} outperforms pattern \ding{173}, highlighting the effectiveness of assigning different weights to sections. Finally, since pattern \ding{174} does not incorporate $\bm{h}_{\text{abstract}}$ (as used in pattern \ding{175}, which only relies on the abstract's overall content as the paper representation), its performance still lags behind the proposed method (pattern \ding{174} $+$ pattern \ding{175}). 
Similarly, pattern \ding{175} also falls short of the proposed method. 
This suggests that integrating both sources of information allows the recommendation system to consider both the overall content of papers and the specific sections that users are particularly interested in (achieved by assigning weights to different sections).
This design aligns with users information seeking behaviors and thus improves recommendation performance. 


\begin{table}[htbp]
\centering
\caption{Ablation study on the impact of different model components in paper recommendation.}
\renewcommand{\arraystretch}{1.4} 
\setlength{\tabcolsep}{8pt} 
\resizebox{\linewidth}{!}{ 
\begin{tabular}{lccccc}
\toprule
\textbf{Pattern} & \textbf{Method} & \multicolumn{4}{c}{\textbf{Metric}} \\
\cmidrule(lr){3-6}
& & recall@5 & precision@5 & MAP & MRR \\
\midrule
& Proposed method  & \textbf{0.8125} & \textbf{0.4591} & \textbf{0.8081} & \textbf{0.8860} \\
\midrule
\ding{172} & Classification model  & 0.7040 & 0.4001 & 0.6939 & 0.8066 \\
\ding{173} & \ding{172} + Title  & 0.7594 &  0.4302 & 0.7522 & 0.8506 \\
\ding{174} & \ding{173} + Attention model  & 0.7727 & 0.4357 & 0.7614 & 0.8523 \\
\ding{175} & $\bm{h}_{\text{abstract}}$  & 0.7881 & 0.4463 & 0.7829 & 0.8693 \\
\bottomrule
\end{tabular}
}
\label{tabl:Ablation}
\end{table}

\subsection{Weight Distribution}
As mentioned in section \ref{sec:sim},  when combining $\bm{h}_{\text{sections}}$ and $\bm{h}_{\text{abstract}}$, we assume that these two components make distinct contributions to the paper's representation. 
Therefore, we introduce the hyperparameter $\alpha$ and 1-$\alpha$, which assign different weights to $\bm{h}_{\text{sections}}$ and $\bm{h}_{\text{abstract}}$, as shown in \eqref{eq:h_paper}. 
Table \ref{tabl:weights} presents the experimental results under various settings of $\alpha$ and 1-$\alpha$. 
Specifically, we vary $\alpha$ from 0.0 to 1.0 in increments of 0.1 to evaluate the impact of different weighting schemes. 

From these results, we observe two important points: (1) when $\alpha$ ranges from 0.1 to 0.5, the combination of $\bm{h}_{\text{sections}}$ and $\bm{h}_{\text{abstract}}$ outperforms each individual component (i.e., setting $\alpha = 0$ or $\alpha  = 1$), and (2) when $\alpha$ is set to 0.3, the model achieves its best performance across multiple evaluation metrics. 
Moreover, when $\alpha$ ranges from 0.7 to 0.9, the performance is worse than using $\bm{h}_{\text{abstract}}$ individually. 
This indicates that appropriate weighting of $\bm{h}_{\text{sections}}$ and $\bm{h}_{\text{abstract}}$ is crucial. 

As noted above, although the model achieves its best performance when $\alpha = 0.3$, we observe that, for $\alpha = 0.4$ or $\alpha = 0.5$, the performance does not decrease significantly across all metrics. 
Moreover, removing $\bm{h}_{\text{sections}}$ ($\alpha = 0$) leads to a significant performance drop (e.g., a decrease in recall@5 or MAP). 
These results suggest that $\bm{h}_{\text{sections}}$ is essential for accurate recommendations, as it contains weighted section-level information.

\begin{table}[htbp]
\caption{Comparison of different weight settings for $\bm{h}_{\text{sections}}$ and $\bm{h}_{\text{abstract}}$.}
\centering
\renewcommand{\arraystretch}{1.4} 
\setlength{\tabcolsep}{8pt} 
\begin{tabular}{cccccc}
\toprule
\multicolumn{2}{c}{\textbf{Weight}} & \multicolumn{4}{c}{\textbf{Metric}} \\
\cmidrule(lr){1-2} \cmidrule(lr){3-6}
$\alpha$ & 1-$\alpha$ & recall@5 & precision@5 & MAP & MRR \\
\midrule
1.0 & 0.0 & 0.7727 & 0.4357 & 0.7614 & 0.8523  \\
0.9 & 0.1 & 0.7668 & 0.4319 & 0.7563 & 0.8503  \\ 
0.8 & 0.2 & 0.7766 & 0.4377 & 0.7638 & 0.8546  \\
0.7 & 0.3 & 0.7847 & 0.4446 & 0.7728 & 0.8573  \\
0.6 & 0.4 & 0.7985 & 0.4438 & 0.7864 & 0.8685  \\
0.5 & 0.5 & 0.8072 & 0.4559 & 0.8010 & 0.8798 \\
0.4 & 0.6 & \textbf{0.8125} & 0.4579 & 0.8069 & 0.8846 \\
0.3 & 0.7 & \textbf{0.8125} & \textbf{0.4591} & \textbf{0.8081} & 0.8860\\
0.2 & 0.8 & \textbf{0.8125} & 0.4586 & 0.8080 & \textbf{0.8874}\\
0.1 & 0.9 & 0.8068 & 0.4558 & 0.8016 & 0.8817\\
0.0 & 1.0 & 0.7881 & 0.4463 & 0.7829 & 0.8693  \\
\bottomrule
\end{tabular}
\label{tabl:weights}
\end{table}


\subsection{Effect of Non-linear Transformation}\label{sec:nonlinear}
As mentioned in section \ref{sec:mlp}, we apply a non-linear transformation to $\bm{h}_{\text{paper}}$, generating $\bm{z}_{\text{paper}}$, which is then used to compute the loss. 
This transformation aim to enhance the quality of $\bm{h}_{\text{paper}}$ (the representation of paper). 
To validate this hypothesis, we conducted experiments, and the results are presented in Table \ref{tabl:nolinear}.

As shown in the top two rows of Table \ref{tabl:nolinear}, we observe that using $\bm{z}_{\text{paper}}$ for loss computation improves performance across all four metrics. 
For instance, recall@5 increases from 0.8049 to 0.8125 when compared to directly computing the loss with $\bm{h}_{\text{paper}}$.
This finding confirms that the non-linear transformation enhances the representation quality of $\bm{h}_{\text{paper}}$.

However, when $\bm{z}_{\text{paper}}$ itself is used as the paper representation on the recommendation task, we observe a decrease in performance. 
This suggests that although the non-linear transformation is beneficial for representation learning, $\bm{z}_{\text{paper}}$ may not be the best choice for directly representing papers in the recommendation stage.


\begin{table}[htbp]
\centering
\caption{Performance comparison of different representations with and without non-linear transformation.}
\renewcommand{\arraystretch}{1.4} 
\setlength{\tabcolsep}{8pt} 
\resizebox{\linewidth}{!}{ 
\begin{tabular}{lccccc}
\toprule
\textbf{Representation} & \textbf{Non-linear} & \multicolumn{4}{c}{\textbf{Metric}} \\
\cmidrule(lr){3-6}
& & recall@5 & precision@5 & MAP & MRR \\
\midrule
$\bm{h}_{\text{paper}}$ & \cmark  & \textbf{0.8125} &  \textbf{0.4591} & \textbf{0.8081} & \textbf{0.8860} \\
$\bm{h}_{\text{paper}}$ & \xmark  & 0.8049 & 0.4552 & 0.8017 & 0.8822 \\
$\bm{z}_{\text{paper}}$ & \cmark  & 0.7846 & 0.4428 & 0.7698 & 0.8561 \\
\bottomrule
\end{tabular}
}
\label{tabl:nolinear}
\end{table}

\subsection{A Case Study of Recommendation}
\begin{figure*}[htb!]
\centerline{\includegraphics[width=\textwidth]{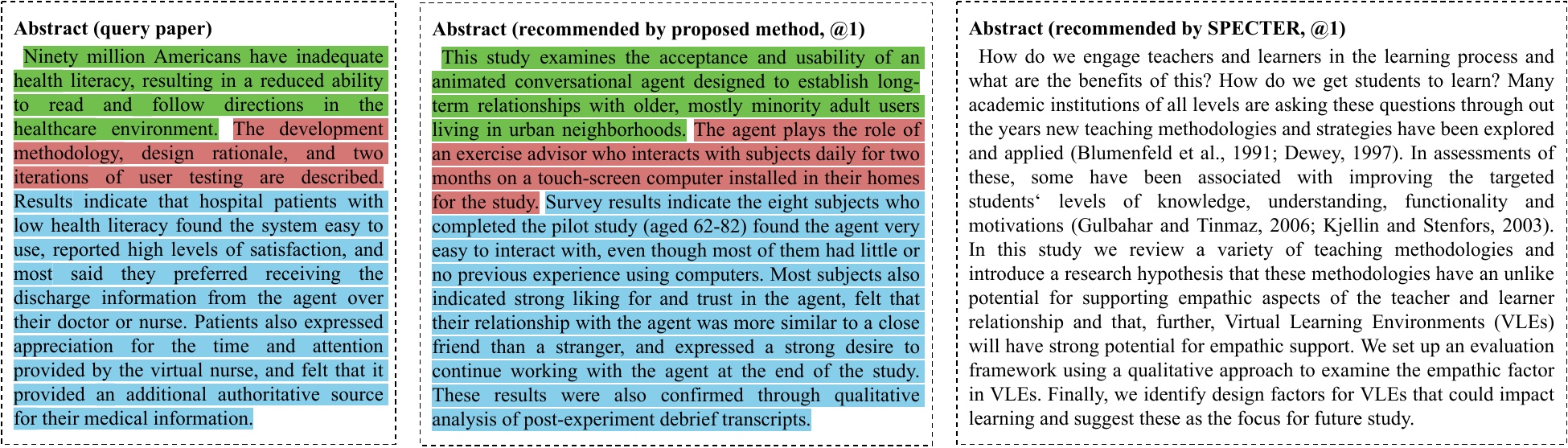}}
\caption{An example of the first recommendation made by our method and SPECTER. The query paper is shown on the left. The paper in the center is recommended by the proposed method and is a relevant paper. 
The paper on the right is recommended by SPECTER and is an irrelevant paper. 
Green indicates the background section, red represents the method section, and blue highlights the results section.}
\label{fig:case_study}
\end{figure*}

We present a case study to illustrate recommendations generated by the proposed method and the best baseline method, SPECTER. 
Fig. \ref{fig:case_study} shows the query paper and the recommendation provided by these two methods. 
We observe that using the proposed method correctly ranks the relevant paper (positive sample) at the first position in the ranked recommendation list, whereas SPECTER instead places an irrelevant paper (negative sample) there. 
Since SPECTER relies only on the overall content of the paper without considering section-level information or user needs, 
it is more likely to recommend an irrelevant paper. 

Because our approach consider users' information seeking behaviors, the proposed method learns a new representation by combining each paper's overall content with three specific sections which are assigned different weights. 
Using the classification model described in section \ref{sec:classification}, the query paper and the relevant paper (center) are classified into three sections represented by green, red, and blue. 
Moreover, as described in section \ref{sec:attention}, our method automatically assigns different weights to each section and these weights reflect user needs. 
In this case study, the wights of the three sections in the query paper were calculated as $W_{background}$ = 0.151, $W_{method}$ = 0.520, and $W_{results}$ = 0.329, respectively. 
This indicates that the user mainly focuses on the method section, followed by the results, and pay the least attention to the background information. 

Based on the recommended paper shown in the center of Fig. \ref{fig:case_study}, 
we observed that the query and recommended paper are generally similar, as both focus on research related to the application of agents. 
Moreover, both papers discuss user testing in their method sections, and the results sections emphasize the effectiveness of agents. 
Although in the background sections, the query paper targets Americans with limited health literacy, while the recommended paper focuses on elderly individuals in urban communities. 
Since the background section has the lowest weight, this suggests that the user is more concerned with the testing of agents and their effectiveness, rather than the target of the agent. 
This leads to the recommendation of this paper as it aligns more closely with the user's needs.

\section{conclusion}
In this paper, we propose a new model for recommending research papers. 
Our model utilizes two key components from each paper's abstract and title to learn paper representations: (1) three specific sections with assigned weight, and (2) overall content of paper. 
We conduct offline evaluations on the DBLP dataset, and the experimental results demonstrate that our approach outperforms six baseline methods across multiple metrics. 
For future work, we plan to evaluate our method on other larger datasets and explore the incorporation of more diverse types of hard negative samples to further enhance model performance. In addition, while we employ an attention model to automatically estimates weights for each section intended to reflect user needs, user studies will be conducted to validate whether these weights accurately capture users' needs in the real-world.


\end{document}